\documentclass[preprint,showpacs,preprintnumbers,amsmath,amssymb,superscriptaddress]{revtex4-2}

\usepackage{graphicx,graphics}   
\usepackage{graphicx}
\usepackage[utf8x]{inputenc}
\usepackage{color}
\usepackage{subfig}
\usepackage{multirow}
\usepackage{slashed}
\usepackage{ulem}
\usepackage{slashed}

\usepackage{graphicx,amsfonts}
\usepackage{epsfig}

\usepackage{dcolumn}
\usepackage{bm}
\begin{document}
	

\title{Neutrino physics and non-standard interactions
}

\author{V. Pleitez }%
\email{v.pleitez@unesp.br; ORCID: 0000-0001-5279-8438 }
\affiliation{
Instituto  de F\'\i sica Te\'orica--Universidade Estadual Paulista \\
R. Dr. Bento Teobaldo Ferraz 271, Barra Funda\\ S\~ao Paulo - SP, 01140-070,
Brazil
}

	
\date{07/01/2022}
	
\begin{abstract}
	
In most of the proposals for new physics beyond the standard model, neutrinos have new interactions and some of them have phenomenological consequences that, although at present they may seem purely academic, probably will have to be taken into account in the future neutrino experiments.
Here we show that new interactions may imply the  misidentification of the flavour of the neutrinos, and the experimental ability to distinguish neutrinos from antineutrinos. 
\end{abstract}

\pacs{
}
	
	
\maketitle
	
\section{Introduction}
\label{sec:intro}

Neutrino physics has entered a precision era, and thus, it arises the possibility to explore new physics beyond the standard model (BSM), not only through out the neutrino oscillations but also with their interactions with hadrons and leptons. For instance, DUNE will be sensitive to electron neutrinos originated in the next supernova explosion~\cite{DUNE:2020zfm} through the study charged-current (CC) interactions with neutrino energies ranging from few MeVs to higher energies. 

Nutrinos with energies of some GeVs or higher can be used for studies of neutrino oscillation parameters and, mainly for searching new physics i.e., interactions beyond those in the standard model (SM) usually called non-standard interactions (NSI). Among other things, the discover of
new particles, new interactions and symmetries beyond those predicted in the context of the SM is an objective for several neutrino experiments, say DUNE~\cite{Abi:2020kei} and others. NSI implies also implies new matter effects
~\cite{Wolfenstein:1977ue,Mikheyev:1985zog}. For a review see Ref.~\cite{Ohlsson:2012kf,Farzan:2017xzy}.

Experiments with neutrinos use, in order to analyze their data, the three-flavor scheme in the context of SM. It is assumed that we in fact know what is the flavor of a neutrino i.e., what is the linear combination of the neutrinos which are symmetry eigenstates, in terms of their mass eigenstates and also the unambiguous identification of neutrinos and antineutrinos. These assumptions are decisive for observing the  effects of the new physics, if it really exists. 

However, things may be not easy. It happens that, precisely the new physics may confuse the interpretation of the observations, if they are interpreted in the aforementioned paradigm. If we define the neutrino of the flavor $l$ the particle which accompany the antilepton $l^+$ and, as being a linear combination of the mass eigenstates determined by the PMNS matrix, we show that there are different linear combinations that can also be defined as "neutrino of flavor $l$". Moreover, in some models, neutrinos accompany leptons and not anti lepton, hence implying a confusion between neutrino and antineutrino in experimental data.

In fact, in Ref.~\cite{Grossman:1995wx} it was noted that new charged current interactions imply that there are different bases for neutrinos and not only the usual symmetry and mass eigenstates and that the identification of the neutrinos cannot be done in a model independent way. Moreover, new interactions may affect also the production and detection of neutrinos modifying the observed event rate and the flavor composition, compared to the SM~\cite{Falkowski:2021bkq}.

Some possibilities for new physics at the TeV scale are the following:
i) Multi-Higgs extension of the SM~\cite{Maniatis:2015gma,Gomez-Bock:2021uyu},
ii) Left-right (symmetric or not) models~\cite{Chavez:2019yal,Diaz:2020pwh},
iii) 3-3-1 models~\cite{Singer:1980sw,Pisano:1991ee,Foot:1992rh,Frampton:1992wt},
iv)  A combination of the previous ones,
v) Grand unification~\cite{Croon:2019kpe},
vi) Supersymmetric version of one of the previous ones~\cite{Montero:2000ng,Rodriguez:2009cd,Babu:2020ncc}, vii) None of the previous ones.
All these possiblities, including the last one, imply the existence of new neutrino interactions.

Here we will show that in fact, in most of the SM extensions it is not possible to define uniquely the flavor of a neutrino and, sometimes we do not distinguish if it is a neutrino or an anti-neutrino.
We give some examples of new charged currents in which the above situation happens.  

The outline of this paper is as follows. In the next section we consider the multi-Higgs extension of the SM; in Sec.~\ref{sec:vectors} we consider interaction of neutrinos with new vector bosons. Examples of effective interactions are given in Sec.~\ref{sec:effectint}. Some consquences of the new interactions are considered in Sec.~\ref{sec:consequences}. Our conclusions are given in Sec.~\ref{sec:con}.

\section{Multi-Higgs doublet models}
\label{sec:escalares}

Let us consider the SM with several scalar doublets. However, our arguments are independent of the type of representation of the scalars.
The representation content is as follows: left-handed leptons in doublets of the SM gauge symmetry $SU(2)_L\otimes U(1)_Y$, $\psi_{L}=(\nu^\prime_{lL}\,l^\prime_{L})^T\sim(\textbf{2},-1)$, and singlets $l^\prime_{R}\sim(\mathbf{1},-2)$, with $l^\prime=e^\prime,\mu^\prime,\tau^\prime$; and also right-handed neutrinos $\nu^\prime_{lR}\sim (\mathbf{1},0)$ and, similarly for quarks. Finally, scalar doublets $H_n=(\phi^+_n\,\phi^0_n)^T\sim(\mathbf{2},+1),\,n=1,2,\cdots$. At this point all states are in the symmetry bases (we omit flavor indices). Teh Yukawa interactions are given by
\begin{eqnarray}
-\mathcal{L}_{Yl}=\sum_n\overline{\psi_{lL}}[G^n_ll^\prime_{R}H_n+G^n_\nu\nu^\prime_{lR}\tilde{H}_n]+H.c.,
\label{higgslnsi}
\end{eqnarray}
where $\tilde{H}^n=i\tau_2H^{n*}$, being $\tau_2$ the usual Pauli matrix. Just for the sake of simplicity, we assume Dirac neutrinos. Mass eigenstates (umprimed fields) are related to the symmetry eigenstates through unitary transformations $l^\prime_L=U^l_Ll_L$, $l^\prime_R=U^l_Rl_R$,  
$\nu^\prime_L=U^\nu_L\nu_L$, $\nu^\prime_R=U^\nu_R\nu_R$ and
\begin{equation}
V^{l\dagger}_LM^lV^l_R=\hat{M}^l,\quad V^{\nu\dagger}_LM^\nu V^\nu_R=\hat{M}^\nu,
\label{massm}
\end{equation} 
where $\hat{M}^{l,\nu}$ are the diagonal mass matrices. Our arguments are valid independently of the neutrino nature (Majorana or Dirac). 

In this model, the lepton-neutral scalar interactions are given by (leptons are in the mass eigenstates basis)
\begin{equation}
-\mathcal{L}^{nc}_{Yl}=\sum_n[
\overline{l_{L}}\mathbf{(A^n_l)}l_{R}\phi^0_n+
\overline{\nu_{L}}\mathbf{(A^n_\nu)}\nu_{R}\phi^{*0}_n+H.c.]
\label{higgsnc}
\end{equation}
where we have omitted generation indices and defined
\begin{equation}
\mathbf{A^n_l}=V^{l\dagger}_LG^n_lV^l_R,\quad  \mathbf{A^n_\nu}=V^{\nu\dagger}_LG^n_\nu V^\nu_R,
\label{higgsdef1}
\end{equation}
and the neutral scalar fields, $\phi^\pm_n,\phi^0_n$, are still in the symmetry bases. The diagonal mass matrices are 
\begin{equation}
\hat{M}^l=V^{l\dagger}_L\left(\sum_nG^n\frac{v_n}{\sqrt2}\right)\,V^l_R,\quad
\hat{M}^\nu=V^{\nu\dagger}_L\left(\sum_nK^n\frac{v_n}{\sqrt2}\right)\,V^\nu_R.
\label{leptonmasses}
\end{equation}

In the mass bases, the charged Yukawa interactions in Eq.~(\ref{higgslnsi}) are written as
\begin{equation}
-\mathcal{L}^{cc}_{Yl}=\sum_n[\bar{\nu}_{lL}(\mathbf{B^n_l})^\dagger l_{R}
\phi^+_n-
\bar{l}_{L}\mathbf{(B^n_\nu)}\nu_{R}\phi^-_n
+H.c.]
\label{higgslcc}
\end{equation}
where
\begin{equation}
\mathbf{B^n_l}=V^{\nu\dagger}_LG^n_lV^l_R,\quad
\mathbf{B^n_\nu}=V^{l\dagger}_LG^n_\nu V^\nu_R.
\label{higgsdef2}
\end{equation}
Notice that both
\begin{equation}
\nu_{lL}=\sum_i(\mathbf{B^n_l})_{li}\nu_{iL},\quad  \nu_{lR}=\sum_i(\mathbf{B^n_\nu})_{li}\nu_{iR}
\label{hnul}
\end{equation}
can be defined as a ``neutrino of flavor $l$", but with different chirality.

The interactions in Eq.~(\ref{higgslcc}) produce transitions as 
\begin{equation} 
e^-_R\to \nu_{eL}+\phi^-_n,
\label{2ndnue}
\end{equation}
instead of the decay in the SM:
\begin{equation}
e^-_L\to \nu_{eL}+W^-,
\label{smnue}
\end{equation} 
and in both cases we can say that it is an ``electron neutrino". However, in Eq.~(\ref{2ndnue}) we have 
\begin{equation} 
\nu_{eL}=\sum_i(\mathbf{B^n_l})_{ei}\nu_{iL}
\label{nhd}
\end{equation}
while in (\ref{smnue}) we have, as usual  
\begin{equation}
\nu_{eL}=\sum_i(\mathbf{V_{PMNS}})_{ei}\nu_{iL} ,\quad \mathbf{V_{PMNS}=
	\mathbf{V_L^{l \dagger}}\mathbf{V^\nu_L}}
\label{sm}
\end{equation}
where $V_{PMNS}$ is the PMNS matrix.

Notice that in general, the PMNS mixing matrix does not appear in the charged interactions with the scalar $\phi^\pm_n$ if $n>1$ and no extra symmetries are added to avoid flavour changing neutral currents. However, even if $n>1$ but if only one doublet generates the lepton masses, the matrices in Eq.~(\ref{higgslcc}) become
\begin{equation}
\mathbf{B^n_l}\to \frac{\sqrt2}{v}\mathbf{V^\dagger_l}\hat{M}^l,\quad \mathbf{B^n_\nu}\to \frac{\sqrt2}{v}\mathbf{V_l}\hat{M}^\nu.
\label{def3}
\end{equation}
In this case there are not NSI, the neutral currents with the scalar $\phi^0$ conserve flavour and are proportional to the lepton masses, $\hat{M}^l$ and $\hat{M}^\nu$. However, as can be seen from Eqs.~(\ref{higgslcc}) and (\ref{higgsdef2}), this is not true in the general case, when two or more Higgs doublets contribute to the lepton masses~\cite{Glashow:1976nt}. Thus, in the latter case only the matrices $\mathbf{B^n_l}$ and $\mathbf{B^n_\nu}$ in Eq.~(\ref{higgslcc}) appears in the charged interactions and there are FCNC in the neutral sector with matrices $\mathbf{A^n_l}$ and $\mathbf{A^n_\nu}$ in Eq.~(\ref{higgsnc}).

In the quark sector the neutral interactions are (also omitting generation  indices)
\begin{eqnarray}
-\mathcal{L}_{Yq}=\overline{Q_{L}}[G^n_qU^\prime_{R}H_n+K^n_d D^\prime_{R}\tilde{H}_n]+H.c.,
\label{higgsqnsi}
\end{eqnarray}

The neutral interactions are (in the mass basis)
\begin{equation}
-\mathcal{L}^{nc}_{Yq}=\sum_n[\bar{U}_L\mathbf{A^n_u} U_R\phi^{0*}_n+
\bar{D}_L\mathbf{A^n_d}D_R\phi^0_n+H.c.],
\label{qnc}
\end{equation}
where 
\begin{equation}
\mathbf{A^n_u}=V^{u\dagger}_LG^n_uV^u_R,\quad  \mathbf{A^n_d}=V^{d\dagger}_LG^n_d V^d_R,
\label{higgsdef4}
\end{equation}
and we have used $U^\prime_L=V^u_LU_L$, $U^\prime_R=V^u_RU_R$, and similarly in the $d$-quark sector. The diagonal mass matrices are:
\begin{equation}
\hat{M}^u= V^{u\dagger}_L\left( \sum_n G^n_u\frac{v_n}{\sqrt2}\right)V^u_R,\quad \hat{M}^d=V^{d\dagger}_L\left(\sum_nG^n_d\frac{v_n}{\sqrt2}\right)  V^d_R,
\label{qmasses}
\end{equation}
and as in the lepton sector, there are FCNCs in the quarks sector if $n>1$ and no extra  symmetries are introduced. It is possible to use the definition of the CKM matrix $\mathbf{V_{CKM}=V^{d\dagger}_L V^u_L}$ to eliminate one of the matrix $V^{U,D}_L$, for instance $V^u_L=V^d_L V_{CKM}$~\cite{Buras:2021rdg}.

The quark charged interactions are (in the mass basis)
\begin{equation}
-\mathcal{L}^{cc}_{Yq}=\sum_n[\bar{D}_L\mathbf{B^n_u}U_R\phi^-_n+\bar{U}_L \mathbf{B^n_d}D_R\phi^+_n+H.c.]
\label{qccs}
\end{equation}
where
\begin{equation}
\mathbf{B^n_u}=V^{D\dagger}_LG^n_uV^U_R,\quad \mathbf{B^n_d}=V^{U\dagger}_LG^n_dV^D_R,
\label{higgsdef3}
\end{equation}
where $U_{L,R}=(u\,c\,\,t)^T_{L,R}$ are mass eigenstates, similarly for $D_{L,R}$. 

In the general case no CKM matrix appears in the charged interactions with the Higgs, ony $\mathbf{B^n_{u,d}}$. Again only when $n=1$ (or $n>1$ but only one Higgs doublet contributes to the quark masses) the CKM and the quark masses appear in Eq.~(\ref{qccs}).
In this case
\begin{equation}
\mathbf{B^n_u}\to \frac{\sqrt2}{v}\mathbf{V_{CKM}}\hat{M^u},\quad \mathbf{B^n_d}\to \frac{\sqrt2}{v}\mathbf{V^\dagger_{CKM}}\hat{M}^d,
\label{qccs3}
\end{equation}
and $\hat{M}^{u,d}$ are the diagonalized mass matrices. In the case in 
Eq.~(\ref{qccs3}), there are not NSI in the quark sector.

Below, we will consider the more general case when $n>1$ and at least two scalar doublets contribute for the quark and lepton masses.

\section{Interactons with vector bosons}
\label{sec:vectors}

Here we use as an example of NSI  the interactions with charged vector bosons, in the minimal 3-3-1 model ~\cite{Pisano:1991ee,Frampton:1992wt,Foot:1992rh}, or in $SU(15)$ GUT~\cite{Frampton:1990hz}.
The interactions with $W^\pm_\mu$ in the lepton sector are as usual involving the PMNS matrix defined as $\mathbf{V_{PMNS}=V^{l\dagger}_LV^\nu_L}$ since in the m331 model, we cannot to begin with the charge lepton mass matrix in a diagonal form \cite{Pleitez:2021abk}. Hence, in the vector charged currents involving neutrinos, there are two matrizes to be determined: besides the $V_{PMNS}$ (appearing in the charged current coupled to $W$), we have the $V_{LR}$ matrices defined below.
Moreover, there are charged currents with  $\nu_L-l^c_L$ and $l_R-(\nu^c)_R$: 
\begin{equation}
\mathcal{L}_{\bar{\nu} l}= i\frac{g}{2\sqrt2}\,\left[\overline{(l^c)_L}\, \mathbf{(V^*_{LR})}\gamma^\mu \nu_{L} -\overline{(\nu^c)_{R}}\,\mathbf{(V_{LR})_{ab}} \gamma^\mu 
l_R \right]V^-_\mu+H.c.
\label{barnul}
\end{equation}
where we have defined
\begin{equation}
\mathbf{V_{LR}}=\mathbf{V^{\nu\dagger}_LV^{l*}_R}.
\label{def}
\end{equation}
Or, we can use in Eq.~(\ref{def}) $\mathbf{V^\nu_L}=V^l_LV_{PMNS}$ and determine the matrices $V^l_L$ and $V^l_R$. 

We can define
\begin{equation}
\nu_{eL}=\sum_i(\mathbf{V^*_{LR}})_{ei}\nu_{iL}.
\label{enuv}
\end{equation}

From Eq.~(\ref{barnul}), we see that transitions as 
\begin{equation} 
\nu_{lL}\to l^+_L+V^-,\quad \nu^c_{lR}\to l^-_R+V^+,
\label{nul331}
\end{equation}
are possible. 

Compare the charged interactions in Eq.~(\ref{barnul}) with that of the SM:
\begin{equation}
\mathcal{L}_{Wl}=\bar{l}_{iL}\gamma^\mu (\mathbf{V_{PMNS}})_{ij}\nu_{jL}W^-+H.c.,
\label{lwi}
\end{equation}
in which the only transitions are 
\begin{equation}
\nu_{lL}\to l^-_L+W^+,\quad (\nu^c)_R\to l^+_R+W^-.	
\label{nulsm}
\end{equation}

It is the interactions with the $W^\pm$ which allow us to define unambiguosly the flavor of the neutrinos: a left-handed (active) electron neutrino is
\begin{equation}
\nu_{lL}=\sum_i(\mathbf{V_{PMNS}})_{li}\nu_{iL},\quad l=e,\mu,\tau,
\label{enu}
\end{equation}
where $\nu_{iL}$ are mass eigenstates. 

The lagrangian terms for interactions among charged gauge bosons and quarks may be written as follows:
\begin{eqnarray}
&& \mathcal{L}_{Wq}=\frac{g}{\sqrt{2}}\bar{D}_{Li}\gamma^\mu
\left(\mathbf{V_{CKM}}\right)_{ij}U_{Lj}W^-_\mu,\quad
\mathcal{L}_{Uj}=\frac{g}{\sqrt{2}}\bar{j}_{Lk}\gamma^\mu\mathbf{(V_L^U)_{kj}}U_{Lj}U^{--}_\mu,\nonumber \\&&
\mathcal{L}_{VJ}=\frac{g}{\sqrt{2}}\bar{J}_{L}\gamma^\mu\mathbf{(V_L^U)_{3j}}U_{Lj}V^{+}_\mu,\quad
\mathcal{L}_{Vj}=-\frac{g}{\sqrt{2}}\bar{j}_{Lk}\gamma^\mu\mathbf{(V_L^D)_{kj}}D_{Lj}V^{-}_\mu,\nonumber \\&&
\mathcal{L}_{UJ}=\frac{g}{\sqrt{2}}\bar{J}_{L}\gamma^\mu\mathbf{(V_L^D)_{3j}}D_{Lj}U^{++}_\mu.
\label{qci}
\end{eqnarray}
Notice that besides the $V_{CKM}$ matrix, which appears in the interactions with $W^\pm_\mu$, the matrices $V^U_L$ and $V^D_L$ appear in the interactions with the vector (bilepton) bosons $V^\pm$ and $U^{\pm\pm}$. $\mathbf{V^{D,U}_L}$ are unitary matrizes used for diagonalizing the respective mass matrizes, $\hat{M}^u=V^{U\dagger}_L M^uV^U_R$, etc. The matrizes $V^{U,D}_R$ survive, separately, in the interactions with the $Z^\prime$~\cite{GomezDumm:1993oxo}.

\section{Effective interactions}
\label{sec:effectint}

Example of effective interactions via charged scalars  from (\ref{higgslcc}) and (\ref{qccs}) are
\begin{equation}
\mathcal{L}_H=\sum_n\left[\frac{1}{m^2_{\phi^+_n}}(\bar{U}_R\mathbf{B^n_u} D_L)(\bar{\nu}_L \mathbf{B^n_l} l_R)+H.c.\right],
\label{eint1}
\end{equation}
wehere we omited flavor indices, and $\mathbf{B^n_l}$ and $\mathbf{B^n_u}$ are defined in Eqs.~(\ref{higgsdef2}) and (\ref{higgsdef3}), respectively.

An example of effective interactions via a vector boson, using Eqs.~(\ref{barnul}) and (\ref{qci}), is 
\begin{equation}
\mathcal{L}_V=\frac{g^2}{2m_{V}m_W}\left(\bar{U}_R \gamma_\mu\mathbf{V_{CKM}} D_L\right)\left(\overline{(l^c)_L}\, \gamma^\mu \mathbf{V^*_{LR}} \nu_L\right)+H.c.
\label{eint2}
\end{equation}
where $\mathbf{V_{LR}}$ is defined in Eq.~(\ref{def}).

\section{Some consequences of new interactions}
\label{sec:consequences}

It seems clear, for what we have discussed above, that the physics beyond the SM may induce confusion in the interpretation of the experimental data: Besides the interactions in Eq.~(\ref{lwi}), 
from which the flavor of a neutrino is defined in the context of the SM, there are other charged interactions in these models.
 We can define different neutrino flavor states:
\begin{eqnarray}
&&\nu^{H_{n}}_{lL}=\sum_{i}\mathbf{(B^n_l)_{li}}\nu_{iL},\quad
\nu^{H_{n}}_{lR}=\sum_{i}\mathbf{(B^n_\nu)_{li}}\nu_{iR}\nonumber \\&&
\nu^V_{lL}=\sum_i\mathbf{(V^\dagger_{LR})_{li}}\nu_{iL}, \quad \nu^W_{lL}=\sum_i\mathbf{(V_{PMNS})_{li}}\nu_{iL},
\label{4types}
\end{eqnarray}
according neutrinos are produced (detected) by the interactions in Eqs.~(\ref{higgslcc}),  (\ref{barnul}) (\ref{lwi}), respectively, $\mathbf{B^n_l}$,  $\mathbf{B^n_\nu}$ and $\mathbf{V_{LR}}$ are defined in Eqs.~(\ref{higgsdef2}) and (\ref{def}), respectively, and the $\mathbf{V_{PMNS}}$ is defined as in Eq.~(\ref{sm}). 
All the states defined in Eq.~(\ref{4types})  are neutrinos that can be considered ``of flavor $l$", since they accompany the charged anti-lepton $l^+$ and, in general
\begin{equation}
	\nu^W_{lL}\not=\nu^V_{lL}\not= \nu^{H_{n}}_{lL}\not= \nu^{H_{n}}_{lR}.
	\label{3tyoes2}
\end{equation}

It is clear that it is not possible to define a neutrino of a given flavor in a model independent way~\cite{Grossman:1995wx}.
Only in models in which the charged leptons are in the mass eigenstates from the very beginning ($\mathbf{V}^l_L=\mathbf{V}^l_R=\mathbf{1}$) we have $\mathbf{V_{LR}=V^\dagger_{PMNS}}$.

Let us look at some consequences of these new interactions of neutrinos.

\subsection{Neutrino oscillation experiments}
\label{subsec:nusosc}

In physics beyond the SM, it is possible that neutrino are produced in the source ($S$) through one interaction and detected in the detector ($D$) with a different interaction. In this case neutrino oscillations have to be consider as in Ref.~\cite{Falkowski:2019kfn},
$$
\textrm{Source}: \quad S\to X_\alpha+\nu,
$$
$$
\textrm{Detection}: \nu+ D\to Y_\beta
$$
where $X_\alpha$ and $Y_\beta$ include hadrons and a charge lepton (or antilepton). Similarly if a antineutrinos are produced and deteted. 

Let us first consider the multi-Higgs extension of the SM with an arbitrary number of doublets. In the most general case, the interaction of leptons with singly charged scalars produces transition as
\begin{equation}
H^-\to l^-_R+(\nu^{H_n}_{lL})^c,\quad (\mathbf{B^n_l})^\dagger 
\label{nus1}
\end{equation}
where $\nu^{H_n}_{lL}$ is defined in Eq.~(\ref{4types}). Notice that, although the particle emitted with the lepton is an (right-handed) electron antineutrino, its components in the mass eigenstates have nothing to do wit the PMNS matrix but with the matrix $\mathbf{B^n_l}$ defined in Eq.~(\ref{hnul}).

It is well known that the  neutrino mass ordering can be normal (NO)  $m_3>m_{1,2}$, or inverted (IO)  $m_{1,2}>m_3$. Present data favoured the NO at least to 1.9$\sigma$~\cite{NOvA:2019cyt}.
However, it was been shown in Ref.~\cite{Capozzi:2019iqn} that the indication in favor of NO when data are interpreted in the standard three-flavor scheme, it does not hold anymore if one assumes the existence of neutral-current nonstandard interactions involving the transitions $e\leftrightarrow \mu$. This happens in all models considered here see, for instance, the interactions in Eq.~(\ref{higgsnc}).

\subsection{Glashow-like resonances} 
\label{subsec:glashowr}

The rate of interaction of $\nu_l$ and $(\nu^c)_R$ with electrons are rather small when compared to interactions with nucleons. This would not be the case for the electron anti-neutrino $(\nu^c_e)_R$ since in this case there is the possibility of the resonant scattering $(\nu^c_e)_R+e^-_R\to W^-\to (\nu^c_e)_R+e^-_R$ or $(\nu^c_\mu)_R+\mu^-_R$~\cite{Glashow:1960zz} and also $(\nu^c_e)_R+e^-_L\to W^-\to \textrm{hadrons}$, when $E_\nu\approx 6$ PeV~\cite{Barger:2014iua} and which are induced by the on-shell production of a $W^-$~\cite{IceCube:2021rpz}.
 
In models with many scalars and new vector interactions there may be more Glashow-like resonances. For instance, in the models that we are consider here 
there are the following possibilities:

\begin{eqnarray}
&&(\nu_{eL})^c+e^-_R\to H^- \to \nu_{lL}+l^-_{R},\quad l=e,\mu,\tau,\nonumber \\&&
(\nu_{eL})^c+e^-_R\to H^- \to \textrm{known hadrons},\nonumber \\&&
\nu_{eL}+e^-_R\to V^- \to \nu_{lL}+l^-_{R},\nonumber \\&&
\nu_{eL}+e^-_R\to V^- \to \textrm{exotic hadrons}
\label{glre}
\end{eqnarray}

At least the resonances involving scalar may be comparable with that of the $W$ is $m_{H^-}=80,90$ GeV~\cite{Dey:2020fbx}. The measurement of the Glashow resonance could make also possible to detect exotic resonances as those pointed out in Ref.~\cite{DeConto:2017aob} predicted in the m331 model.

\subsection{Neutrino or antineutrino?}
\label{subsec:nuantinu}

Let us consider the neutrino-antineutrino confusion that occurs in some models. As we said before, usually it is considered that the detector is able to discriminate between $\nu_e$ and $\nu^c_e$. 
However, in some models it may be difficult to decide if a neutrinos or an antineutrino has been detected. 
Let us illustrate this situation considering the m331 in which there is a second, singly charged, vector boson arises~\cite{Pleitez:2021abk} and interactions as those given in Eq.~(\ref{barnul}). Hence, transitions like those in Eq.~(\ref{nul331}), $V^+\to (\nu^c_{l})_R+ l^+_L$, are possible and,
if interpreted in the SM context, it will be considered an antineutrino, not a neutrino!

Besides, the vertex has a mixing matrix which is different from the usual PMNS matrix: $ V_{LR}\not= U_{PMNS}$. The transition 
\begin{equation} 
(\nu^c)_R\to V^++l^-_R,
\label{vint2}
\end{equation}
also occurs. We see that a complete identification of neutrino or antineutrinos needs a measurement of the helicities of the  charged lepton, or the detection of the vector boson. 

The reaction (inverse muon decay) $\nu_{\mu L}+e^-_L\to \mu^-_L+\nu_{e L}$, has a cross-section that can be predicted with very small uncertainties. It has a neutrino energy threshold of $\approx11$GeV and is used used to constrain the high-energy part of the flux in the NuMI neutrino beam~\cite{Ruterbories:2021myb}. 

It is worth to pointed out that in the m331 model we have the decay
\begin{eqnarray}
&&(\nu^c_\mu)_R+e^-_R\to V^-\to\mu^-_R+(\nu^c_e)_R,\quad t-\textrm{channel}\nonumber \\&&
\label{inmuon}
\end{eqnarray}

Summaryzing: In the context of the SM we have the processes
\begin{eqnarray}
&& \nu_{eL}\to l^-_L+W^+,\quad \textrm{neutrino},\quad (a)\nonumber \\&&
\nu^c_{eR}\to l^+_R+W^-,\quad \textrm{anti-neutrino}\quad (b)
\label{wm331}
\end{eqnarray}
and, in the m331 model, besides the transitins in Eq.~(\ref{wm331}), also occur
\begin{eqnarray}
&& \nu_{eL}\to l^+_L+V^-,\quad \textrm{neutrino!}\quad (a)\nonumber \\&&
\nu^c_{eR}\to l^-_R+V^-,\quad \textrm{anti-neutrino!} \quad (b)
\label{vm331}
\end{eqnarray}

The helicity of the charged lepton allows to distinghis an neutrino with a $W^+$, as in Eq.~(\ref{wm331})(a), or an antineutrino if a $V^-$ is emitted as in Eq.~(\ref{vm331})(b).

Although the confusion neutrino-anti-neutrino may not have a considerable effect in active neutrino oscillations, this is not necessarily the case with rare processes as  neutrino$\leftrightarrow$anti-neutrino oscillations. This sort of oscillations may be viewed as the process 
$W^++l^-_L\to \nu\to (\nu^c)_R\to  l^+_R+W^-$
in which the neutrino, in the intermediate-state,  travels a macroscopic distance~\cite{deGouvea:2002gf}. In this case the intermediate states requires a mass insertion $\nu\to \bar{\nu}$. In the m331 we have $l^-_L+W^+\to\nu_L\to l^+_L +V^-$ (no helicity suppressed). Thus, in principle, both mechanism are distinguishable since the decays of $W$'s are different from those of $V^-$ and in the latter case there is no helicity suppression.

\subsection{Other processes}
\label{subsec:others}




In the m331 model there are processes induced by exotic quarks, the extra neutral $Z^\prime$, and by the vector bosons, $V^-$ and $U^{++}$. The interactions of the singly charged vector boson with leptons are those in Eq.~(\ref{barnul}) while for the interactions of the doubly charged vector with leptons and quarks see Ref.~\cite{Machado:2016jzb}. For instance, the processes 
\begin{widetext}
\begin{eqnarray}
&& \nu_{\mu L}+N\to \mu^+_L+X^-_j(jud)\;\;(a),\;\;\bar{\nu}_{\mu R}+N\to \mu^-_R+Y^-_J(Jdd)\;\; (b),\nonumber \\&&
\stackrel{(-)}{\nu_{a L}}+N\to  l^-_{1L} l^+_{2L} l^-_{3R}+X^-\;\; (c), \;\;
\nu_{\mu L}+e^-_L\to l^-_{1L} l^+_{2L} l^-_{3R} \nu_{a L} \;\;(d),\nonumber \\ &&
\stackrel{(-)}{\nu_a}+N\to \stackrel{(-)}{\nu _\ell} l^+l^-+X \;\;(e),\;\; \mu^-_L+N\to \mu^+_Ll^-l^-+N \;\;(f),
\nonumber \\&&
\stackrel{(-)}{\nu _a}+N\to \stackrel{(-)}{\nu _\ell}+X,\;\;(g),\;\; 
\nu+N\to \mu^-+X\;\;(h),
\label{123}
\end{eqnarray}
\end{widetext}
may be observed in experiments like the PINGU~\cite{Aartsen:2014oha} and ORCA~\cite{Adrian-Martinez:2016fdl}. There are also
new contributions to the leptonic decays, for instance $\mu^+_R\to (\nu_{\mu L})^c+e^-_R+\nu_{eL}$, are allowed. 

Some of the processes above also occur in the standard model. However, even in these cases there are important differences. For instance, in the SM we have $\nu_{\mu L}(\bar{\nu}_{\mu R})+N\to \mu^-_L(\mu^+_R)+\textrm{hadrons}$, where ``hadrons" means the known hadrons and only left-handed currents are involved. In Eq.~(\ref{123}a), the elementary process is $\nu_{\mu L}+d_L\to \mu^+_L+j_L$, and the reaction proceeds because in the m331 model there are  also right-handed currents in the leptonic sector. Moreover, $X_j$ and $Y_J$ denote new hadronic resonances which include exotic quarks, in the example above, a quark $j(J)$ with electric charge $-4/3(5/3)$ (in units of the positron electric charge) is being considered. Reactions involving anti-neutrinos, as in Eq.~(\ref{123}b), occurs via a doubly charged vector boson $U^{++}$ and the resonance involves the exotic quark with electric charge $+5/3$, $J$.
The lightest of these resonance can decay only into the known leptons, for instance, $X_j^-\to V^-+\nu\to l^-_L+\nu_{lL}+\nu$. Since the reactions in Eqs.~(\ref{123}a) and (\ref{123}b) create different hadrons we expect that a charge asymmetry may appear at some energy. This reminds us the high-$y$ anomaly events (recall $y=(E_\nu-E_\mu)/E_\nu\equiv E_h/E_\nu$ where $E_h$ is the hadron energy) which was an excess in $\nu_\mu(\bar{\nu}_\mu)+N\to \mu^-(\mu^+)+X$, with respect to the electroweak standard model,~\cite{Aubert:1974en,Benvenuti:1976ad,Benvenuti:1976nq}. At the time, this was considered as evidence of violation of the charge symmetry induced by, i) new particles, ii) the existence of right-handed currents, or iii) even a breaking of scale invariance. Although this anomaly was not confirmed at neutrinos with energies between 30 and 200 GeV~\cite{Holder:1977en,delAguila:2009bb}, the m331 model predicts that they may exist at some scale of energy.

The reaction in Eq.~(\ref{123}c) again remind us the old effect: the trimuon events $\nu_\mu+N\to3\mu+X$~\cite{Benvenuti:1977zp,Benvenuti:1977fb,Holder:1977gp}. If this reaction had been confirmed it would imply i) new heavy neutral leptons or, ii) new gauge bosons. See Fig.~1 of Ref.~\cite{Barela:2019pmo}.
A purely leptonic process is that in Eq.~(\ref{123}d).
The reaction Eq.~(\ref{123}e) is an example of trident neutrino production which in this model arises as in Refs.~\cite{Altmannshofer:2014pba,Ge:2017poy}, and  the processes with three leptons in Eq.~(\ref{123}f) occurs via a doubly charge boson, say $U^{--}$. Of particular interest could be the neutrino-nucleon scattering via neutral currents in Eq.~(\ref{123}g) and through charged currents in Eq.~(\ref{123}h) at low $Q^2$ as in ~\cite{Zeller:2001hh}. This will provide a measurement of $\sin\theta_X(Q)$ at $Q\simeq$ MeV-GeV range, and it can be verified if at these energies  $s^2_X\approx s^2_W$. The process in Eq.~(\ref{123}h) is well known but surprises may arise studying it in more details.

Moreover, for processes induced by subprocesses  $V^-V^-\to l^-l^{\prime -}$ and $U^{--}\to  l^-l^{\prime -}$, the extra neutral vector boson $Z^\prime$ will also give interesting signature in left-right assymmetry in lepton-lepton colliders~\cite{Montero:2000ch}. 

In the m331 model the interactions of leptons with $Z$ and $Z^\prime$ are universal. However, in the limit in which the couplings of fermions with the $Z$ goes to the SM value, the couplings with $Z^\prime$ are only functions of $s^2_S$. As an example, for neutrinos when $g^\nu_V=g^\nu_A=1/2$, the couplings with the $Z^\prime$ are $f^\nu_V=f^\nu_A=-(\sqrt3/6)(1-4s^2_W)^{1/2}$~\cite{Dias:2006ns}. Hence, it is interesting to search for effects of $Z^\prime$ in $\nu N$~\cite{Cheung:2021tmx} and $Z^\prime\to4\nu$~\cite{Belotsky:2001fb}.

All these processes discussed above have contributions from neutral and charged scalars that in general are not negligible~\cite{Barela:2022sbb}.

\section{Conclusions}
\label{sec:con}

We have shown here, with some concrete examples, that in models beyond the SM, there are different bases for a neutrino of flavor $l$, and that in some models, a neutrino can be confused with an anti-neutrino if an anti-neutrino of flavor $l$ is defined as the one which is emitted with the lepton $l^-$.

Everything we have discussed above can also occur in others well motivated models, for example models with the gauge symmetry~\cite{Senjanovic:2017ldw,Senjanovic:2019moe,Chavez:2019yal,Diaz:2020pwh}. In the most general case, when parity is broken explicitly~\cite{Diaz:2020pwh}, independently if neutrinos are Dirac or Majorna particles $V^L_{PMNS}\not=V^R_{PMNS}$. 
\begin{equation} 
SU(2)_L\otimes SU(2)_R\otimes U(1)_{B-L},\quad g_R(\mu)\not= g_L(\mu),\;\forall \mu.
\label{lrm}
\end{equation}
Hence, if neutrino are Dirac particles we have two independent linear combinations:
\begin{equation}
\nu^{W_L}_{aL}=\sum_i(V^L_{PMNS})_{ai}\nu_{iL},\quad \nu^{W_R}_{aR}=\sum_i(V^R_{PMNS})_{ai}\nu_{iR},\quad
i=1,2,3.
\label{nulnur}
\end{equation}
Notice that, in this case, $\nu^{W_L}_L\not= \nu^{W_R}_R$.
Moreover, there is a new $C\!P$ violating phase between the interactions (in quark and lepton sectors) with $W^\pm_L$ and $W^\pm_R$. The same occurs in 3-3-1 model and the extra singly charged vector boson, $V^\pm$~\cite{Pleitez:2021abk}. 

In summary, if neutrinos have new interactions the definition of the flavor of a neutrino to be model dependent. Even the experimental distinction of what is a neutrino or an antineutrino can be complicated because in some models neutrinos have interactions that fake an antineutrino.

\end{document}